\documentclass[bibyear]{aa}  

\usepackage{graphicx}
\usepackage{color}

\usepackage{natbib}
\usepackage{txfonts}
\defcitealias{PerriLeitner2022}{Perri \& Leitner et al.} 

\begin{document} 

   \title{Filament eruption by multiple reconnections}
   
    \author{Y. Liu\inst{1}
            \and
            G. P. Ruan\inst{1}
             \and
            B. Schmieder\inst{2}\fnmsep \inst{3}\fnmsep \inst{4}
\and
            J. H. Guo\inst{2,6}
                \and
            Y. Chen\inst{1}
            \and
            R. S. Zheng\inst{1}
            \and
            J. T. Su\inst{5}
            \and
            B. Wang\inst{1}
            }

    \institute{Shandong Provincial Key Laboratory of Optical Astronomy and Solar-Terrestrial Environment, and Institute of Space Sciences, Shandong University, Weihai 264209, China\\
            \email{rgp@sdu.edu.cn}
            \and
            Centre for mathematical Plasma Astrophysics, Dept.of Mathematics, KU Leuven 3001,  Leuven, Belgium
            \and
            Observatoire de Paris, LESIA, Universit\'e PSL, CNRS, Sorbonne Universit\'e, Universit\'e de Paris, 5 place Jules Janssen, F-92190 Meudon, France
            \and
            University of Glasgow,  School of Physics and Astronomy,  Glasgow G128QQ, Scotland, UK
             \and
            Key Laboratory of Solar Activity, National Astronomical Observatories, Chinese Academy of Sciences, Beijing 100012, China
            \and 
            School of Astronomy and Space Science and Key Laboratory of Modern Astronomy and Astrophysics, Nanjing University, Nanjing 210023, China 
             }

 
  \abstract
   {Filament eruption is a common phenomenon in solar activity, but the triggering mechanism is not well understood. 
   }
   { We focus our study on a filament eruption located in a complex nest of three active regions close to a coronal hole.
   }
   {The filament eruption is observed  at multiple wavelengths: by the Global Oscillation Network Group (GONG), the Solar Terrestrial Relations Observatory (STEREO), the Solar Upper Transition Region Imager (SUTRI), and the Atmospheric Imaging Assembly (AIA) and Helioseismic and Magnetic Imager (HMI) on board  the Solar Dynamic Observatory (SDO).  Thanks to high-temporal-resolution observations, we were able to analyze the evolution of the fine structure of the filament in detail. The filament changes direction during the eruption, which is followed by a halo coronal mass ejection detected by  the Large Angle Spectrometric Coronagraph (LASCO) on board the Solar and Heliospheric Observatory ((SOHO).
   A Type III  radio burst was also registered at the time of the eruption. To investigate the  process of the eruption, we analyzed the magnetic topology of the filament region adopting a nonlinear force-free-field (NLFFF) extrapolation  method and   the polytropic global magnetohydrodynamic (MHD) modeling. We  modeled the filament by embedding a twisted flux rope with the regularized Biot-Savart Laws (RBSL) method in the ambient magnetic field.
   }
   {The extrapolation results show that magnetic reconnection occurs in a fan-spine configuration resulting in a circular flare ribbon. The global modeling of the corona demonstrates that there was an interaction between  the filament and open field lines, causing a deflection  of the filament in the direction of the observed CME eruption 
   and   dimming  area. 
   }
   {The modeling supports the following scenario: magnetic reconnection not only occurs with the filament itself (the flux rope) but also with the  background magnetic field lines and open field lines of the  coronal hole located to the east of the flux rope. This multiwavelength analysis indicates that the filament undergoes multiple magnetic reconnections on small and large scales with a drifting of the flux rope. }

   
\maketitle

%

\section{Introduction}
Solar filaments consist of relatively cold, dense plasma and are usually located above the polarity inversion lines \citep[PILs;][]{Babcock1955,Zirker1989}. When observed with H$\alpha$ spectral lines on the solar disk, filaments appear as long dark structures, showing absorption characteristics. When they move to the limb of the Sun, they appear as bright emission features relative to the dark background, which is why they are often referred to as prominences \citep{Ruan2018,Ruan2019a}. Filament eruptions are often associated with other forms of solar activity, such as solar flares, jets, and coronal mass ejections (CMEs). 

The fine structure of the filaments
can be analyzed thanks to high-resolution observations. Such images show that filaments are made of many  parallel strands that are highly dynamic,  showing a  variety of flows. \citet{Zirker1998} observed counter-streaming along a prominence spine and barbs in H$\alpha$ and detected these flows with speeds in the range of 5–20 km/s. For active region filaments, during reconnection between strands,  H$\alpha$ blobs can be observed  with similar velocities \citep{Deng2002}. 
During  filament activity, counter-streaming in hotter plasma is also detected with velocities in the plane of the sky of between 70 and 100 km/s in   193 \AA{} \citep{Alexander2013}. The flows of filament material must move along the magnetic field lines under the magnetic freezing effect. In this case, filament strands are bound by magnetic field lines, and so the direction of strands can often be used as a tracer of local magnetic field lines. 

There are three general types of filament eruptions. The first type is failed eruptions \citep{Liu2009,Kumar2011, Shen_yd2011,Kumar2022,Joshi2022}, where the filament cannot break free from the solar bond without accompanying CMEs. The second is partial eruptions \citep{Tripathi2009,Gibson2006,Joshi2014,Schmieder2014,Bi2015,Cheng2018,Monga2021,Zhang_yj2022,Yang_jy2023}, in which part of the filament falls back to the surface of the Sun while some manages to escape. The third is successful eruptions {\citep{Gilbert2007,Schrijver2008,Shen2012,Schmieder2013,Zhang_yy2022,Ruan2014,Ruan2015}, in which the filaments all explode away from the Sun and accompany the CMEs.

When the equilibrium state of a filament is destroyed by one or more factors, such as shear motion of the photosphere magnetic field \citep{Sakai1992}, magnetic appearence and cancelation \citep{Martin1986}, large-scale changes in the magnetic field structure of the corona \citep{Low1990}, and so on,  its magnetic field configuration can change, and this can trigger a filament eruption. In recent years, researchers have mainly divided the triggering mechanisms of the outbreak of filaments into two categories. One of these is magnetic reconnection \citep{Su2007,Schmieder2012,Schmieder2013,Chen2014,Schmieder2016,Zheng2017,Li2018,Chen2018,Hou2019,Ruan2019b,Yan2020,Chandra2021,Guo_yl2021,Hu2022,Koleva2022,Sun2023,Yang_lp2023,Xue2023,Hou2023}. The ideal magnetohydrodynamic instabilities, such as kink instability \citep{Gold1960,Hood1981} and torus instability \citep{Kliem2006}, are also considered to be a plausible explanation for filament eruptions \citep{Xu2020}.

\citet{Song2018} detected  a white-light enhancement during a flare that may be caused by the internal reconnection of the flux rope  or by the reconnection between the flux rope and the overlying  magnetic field. \citet{Zheng2019} studied a confined partial eruption involving
at least two magnetic reconnections. \citet{Zhou2021} observed a filament eruption going through internal and external magnetic reconnection, which triggered another filament eruption. Interestingly, the latter also experienced  internal and external magnetic reconnection.  \citet{Zhang_yj2022} demonstrated that another scenario of partial eruption is possible, in which the higher filament fibrils reconnect with closed loops and therefore fail to escape; the lower filament fibrils reconnected with open field lines and developed into a CME. \citet{Zhang_yy2022} reported a ``double-decker'' filament formation, where the upper filament erupted by internal magnetic reconnection and magnetic field motion.  \citet{Dai2022} observed direct evidence of the filament threads reconnecting with the ambient loops, resulting in two eruptions accompanying a filament split.  \citet{Li2023} studied a mini-filament eruption and found evidence of external reconnection between the filament and ambient loops. \citet{Guo2023b} studied a filament eruption that exhibited significant lateral drift while it rose up; external magnetic reconnection caused the flux rope footpoints to migrate, causing the subsequent CME to deviate from the filament. 

It is widely believed that corona dimming is caused by the loss of plasma during CMEs \citep{Tian2012,Jin2022}. \citet{Sterling1997} first confirmed twin dimming in the X-ray band. \citet{Zarro1999} analyzed the same CME event and found that dimming also occurred in the EUV band in the same region as the X-ray band dimming, with similar morphology, and this dimming was observed in different EUV bands of SOHO/EIT. The dimming morphology found by \citet{Jiang2003} in the H$\alpha$ image is the same as this twin dimming. Specifically, dimming areas are believed to correspond to the footpoint of the erupting flux rope. Coronal dimmings take place in the impulsive phase of the flare, which is consistent with the acceleration phase of the corresponding CME \citep{Cheng2016}. The typical evolution of coronal dimming is characterized by a sharp rise followed by a slow recovery by magnetic reconnection \citep{Attrill2006,Reinard2008}. Remote dimming may be related to the circular-ribbon flare connected to the active region by large-scale coronal loops \citep{Zhang2020}. 

Some filaments and CMEs do not propagate in the radial direction of the Sun during ejection. \citet{Gopalswamy2004,Gopalswamy2009} reported that coronal holes are responsible for the deflection of CMEs from the Sun--Earth line. \citet{Shen2011} reported the dynamical evolution of a CME, and their observations showed a CME that is significantly deflected by about $30 ^{\circ}$ from the low-latitude region at the beginning, after which the CME propagates in a radial direction. These authors also found that the early deflection of the CME may be due to the inhomogeneous distribution of the background magnetic field energy density. \citet{Gui2011} used a model to statistically analyze the deflections of ten CME events observed by STEREO, and found that the deflections were the same as the gradient of the magnetic energy density in terms of strength and direction. \citet{Yang2015} studied a filament eruption that was first guided by the open field and was then deflected by a nearby coronal hole before being reconnected with the open field in the opposite direction of the remote coronal hole. \citet{Yang2018} reported a $90 ^{\circ}$ deflection of a CME with respect to its initial propagation direction due to the effect of the open magnetic field of a coronal hole. In addition to the deflection of CMEs related to the nearby coronal holes and the gradient of the magnetic energy density, strong magnetic structures in the vicinity can also cause deflections of CMEs \citep{Jiang2007,Bi2011,Bi2013}.

In this paper, we present multiwavelength observations  of a filament eruption obtained from  the Global Oscillation Network Group (GONG), the Solar Terrestrial Relations Observatory (STEREO), the Solar Upper Transition Region Imager (SUTRI), and the Atmospheric Imaging Assembly (AIA) and Helioseismic and Magnetic Imager (HMI)  on board  the Solar Dynamic Observatory (SDO). We describe how we analyzed the dynamics of the filament eruption process in detail, and then we summarize our results and  discuss the possible triggering mechanism of the eruption.

\section{Data}

We examined observations of  the filament eruption obtained by multiple instruments. All of the data we used were obtained in the time range of 12:00-16:00 UT on 4 October 2022. 
\subsection{SDO}    
The AIA \citep{Lemen2012} on board the SDO \citep{Pesnell2012} provides simultaneous high-resolution full-disk images of the chromosphere, the transition region, and the corona with a pixel resolution of $0.^{\prime\prime}6$ and a temporal resolution of 12s. The datasets we used contain seven extreme-ultraviolet (EUV) wavelengths. 
The HMI \citep{Schou2012} on board the SDO provides four main observables. We downloaded line-of-sight magnetograms with a cadence of 45s and a pixel size of $0.^{\prime\prime}5$. 
\subsection{STEREO}    
 STEREO \citep{Kaiser2008} is made up of a pair of twin satellites, namely STEREO-A (Ahead) and STEREO-B (Behind). The EUV imager (EUVI) \citep{Wuelser2004} of the Sun--Earth Connection Coronal and Heliospheric Image (SECCHI) on STEREO-A provides four wavelength channels. We downloaded 195 \AA{} data ---with the main contributing ion being Fe XII--- with a cadence of 2.5 min and a pixel size of $1.^{\prime\prime}58$.
\subsection{SUTRI} 
SUTRI \citep{Bai2023} on board the Space Advanced Technology demonstration satellite (SATech-01) provides solar-disk images at 46.5 nm. The Ne VII 46.5 nm line forms in a region of about 0.5 MK in the solar atmosphere (located in the high transition zone). We obtained the data with a cadence of 30s and a spatial resolution of $1.^{\prime\prime}23$ per pixel. The Ne VII 46.5
nm line  is in a key region connecting the lower atmosphere to the corona. This is the first time that China has conducted solar transition zone exploration.
\subsection{GONG}  
GONG \citep{Harvey1996} provides H$\alpha$ data. The spatial and temporal resolutions are about 1$^{\prime\prime}$ per pixel and 1 minute, respectively.

\section{Observations}

We focus on the partial eruption of a filament that occurred on 4 October 2022. The filament in question was a large-scale filament located in the southern hemisphere (x = -200, +200 arc sec, y= -600,-450 arc sec) surrounded by three active regions, namely NOAA 13114, 13115, and 13117, parts of which show nonradial eruption. Although no X-ray flares or H$\alpha$ flares were observed during this eruption, there were two obvious flare ribbons and post-flare loops accompanied by a halo CME. The position  of the  eastern part of the dark filament is shown in Figure~\ref{position} marked by a cyan contour. The filament can be seen in both EUV and H$\alpha$ bands. This eruptive filament lies on the polarity inversion lines (PILs), and consists of a large region of mixed polarity; it was the southern part of it that erupted. 
On the eastern side  of the filament, a large coronal hole was present.  In order to study the dynamics of its eruption in detail, we obtained simultaneous observations made by the SDO, STEREO, SUTRI, and GONG instruments in order to observe the filament eruption in the different layers of the Sun's atmosphere. We summarize information regarding the major activities of the eruption  in  Table \ref{tab:1}.

\begin{table*}[htbp]
\centering
\caption{Observed information regarding the major activities of the eruption}
\label{tab:1}
\resizebox{\linewidth}{!}{
\begin{tabular}{clll}
\hline\hline\noalign{\smallskip}
Time (UT) & Observation & Instrument & Wavelength/Frequency  \\
\noalign{\smallskip}\hline\noalign{\smallskip}
13:20:05-14:01:05 & Filament eruption      & SUTRI, AIA, STEREO, GONG & 465\AA{}, 304\AA{}, 171\AA{}, 211\AA{}, 335\AA{}, 131\AA{}, 94\AA{}, 195\AA{}, 6562.8\AA{}   \\ 
14:00:05-15:28:05 & Filament deflection    & AIA & 304\AA{}  \\ 
13:58:05-16:00:00 & Remote dimming         & SUTRI, AIA, STEREO & 465\AA{}, 304\AA{}, 171\AA{}, 211\AA{}, 335\AA{}, 131\AA{}, 94\AA{}, 195\AA{}   \\ 
13:40:06-14:20:06 & Twin dimming           & AIA & 193\AA{}, 211\AA{}  \\ 
14:10:00-14:10:30 & Type \uppercase\expandafter{\romannumeral3} radio burst    & HUMAIN   & 45-450 MHz  \\ 
14:12:05-15:42:05 & CME                    & C2, C3  & \\ 
\noalign{\smallskip}\hline
\end{tabular}
}
\end{table*}

In the multiwavelength   movies provided by AIA, we see that the filament initially erupts in the northeast direction, and then goes through some process that causes it to erupt towards the southwest. The filament begins to lift up, as can be observed  in all the EUV wavelengths around 13:25 UT. As the filament rises, two bright flare ribbons appear below it. These two bright ribbons are nearly parallel and gradually move away from each other with time, on both sides of the PIL. It is worth noting that one of the flare ribbons expands northward along a circular trajectory and shows a remote brightening at the end (see Figure~\ref{eruption}); this flare ribbon extends to the south of  the active region AR 13117. The post-flare loops are also visible in the EUV bands. All of these activities may be the result of magnetic reconnection. 

The filament keeps rising in a northeast direction and begins to erupt around 14:00 UT (see white and red arrows in Figure~\ref{eruption}).  From the large field of view of the 304 \AA{} images shown in Figure~\ref{eruption}, the eruption can be clearly divided into two stages.  In the first stage, the direction of elevation  is northeast. The nonradial direction of the filament's initial eruption is also noteworthy. However, as the filament reaches a certain height, in the second stage, the direction is distinctly different from that of the previous stage. The direction of the eruption is now southwest. We would like to know what causes this deflection in direction. The entire eruption process is shown in Figure \ref{eruption} and the positions of several slices are plotted on it.  A slice diagram of the filament eruption is shown in Figure~\ref{304slice}. 

The slice diagram in panel (a) of Figure~\ref{304slice} shows an obvious change in the direction of the filament during the eruption. S1 is a line slice along the direction of filament eruption (initial stage), S2 and S3 are arc slices along the dimming region propagation and filament deflection, respectively. Along the direction of S1, the filament rises slowly at first and then accelerates, and its velocity after deflection is followed by a speed of 101 km/s.  This diagram demonstrates the sweeping motion of the filament in a southwestern direction along S3 with a speed of 123 km/s.

During the eruption, coronal dimming is observed in the northeast and propagates anticlockwise in the EUV bands. Meanwhile, we also observe a bright flare ribbon extending to the dimming area. There is a temporal correlation between the dimming region and the subsequent deflection.  Figure~\ref{dimming} shows time--distance diagrams of four EUV bands along S2, which is the direction in which the dimming area propagates. The remote brightening and remote dimming   region (S2) are represented by black and white arrows, respectively. The dimming comes after the remote brightening, with a delay of about 25 minutes. As the dimming propagates,  a continuous brightening appears in front of it in the 304 \AA{} images.  

In the movie of 193 \AA{},  the ends of two parallel flare ribbons bend into hook shapes ---marked by yellow ovals  in Figure~\ref{flux}--- during the eruption, where  the twin dimming occurs. The areas of the hooks shrink gradually with time,  similarly to the process described by \citet{Aulanier2019}. The twin dimming corresponds to the two footpoints of the magnetic flux rope that subsequently erupts.  We selected the  twin dimming locations, which we denote LD and RD }(white boxes in Figure~\ref{flux} panel (a)),   and calculated the magnetic flux for the two individual dimmings, as shown in   panels
(b) and (c) of Figure~\ref{flux}. These two dimming regions correspond to the two footpoints of the filament.  The left side of the filament is rooted in negative polarity, and the negative magnetic flux at the left footpoint gradually decreases, while the positive flux  increases before the eruption. For the right footpoint, it is rooted in positive polarity.  Positive flux first increases and then decreases for a period of time,   while negative flux decreases by a small amount before the eruption  but increases after the eruption. This indicates that magnetic emergence and magnetic cancelation occur at both footpoints before the eruption of the filament. Magnetic activities at the footpoints make the filament increasingly unstable, setting the stage for the subsequent eruption. 
During the eruption, the two ribbons move away from each other, and  as the filament changes direction, the  left leg of the flux rope approaches the coronal hole  (indicated as CH marked in white  in panel
(a) of Figure \ref{flux}).
The filament   field lines could interact with the open field lines of the CH according to the movie (movie\_bd.mp4).

We obtained the radio dynamic spectra associated with this filament  eruption from observations by HUMAIN\footnote{https://www.sidc.be/humain/callisto\_burst\_archives} of the Royal Observatory of Belgium, which is part of the network of Compound Astronomical Low frequency Low cost Instrument for Spectroscopy and Transportable Observatory (CALLISTO). There is a type \uppercase\expandafter{\romannumeral3} radio burst around 14:10 UT shown in Figure~\ref{radio}, which indicates the interaction with open field lines, which could correspond to the coronal hole. 

During the filament eruption, a halo CME was observed by the Large Angle Spectrometric Coronagraph  \citep[LASCO;][]{Brueckner1995} on board SOHO \citep{Domingo1995}. The CME started at 14:12:05 UT with a linear fit velocity of 926 km/s (see green arrows in Figure~\ref{cme}) provided by the SOHO LASCO CME catalog\footnote{https://cdaw.gsfc.nasa.gov/CME\_list/}. The direction of the CME is roughly radial. In the movie, the CME shows close correspondence with the eruption.

\section{Three-dimensional magnetic field modeling}

To investigate potential reconnection geometries during the eruption, we reconstructed 3D coronal magnetic fields involved in the filament eruption. Nonlinear force-free-field (NLFFF) extrapolation ---concentrating on the magnetic configuration  of the active region---  and polytropic global coronal MHD modeling are adopted in this paper.  Importantly, we note that it is the typical magnetic topology that is responsible for magnetic reconnection processes taking place in the low and high corona. 

The NLFFF extrapolation is achieved using the magneto-friction relaxation module in the MPI-AMRVAC framework  \citep{Guo2016a, Guo2016b, Xia2018}. In this model, the magnetic field is totally dominated by the magnetic-induced equation with a  velocity nearly  proportional to the local Lorentz force, such that the converged state approaches a force-free state. The implementation of the NLFFF extrapolation includes the preprocessing for the bottom boundary condition and computation of coronal magnetic fields, as follows.

First, we carried out pre-processing of the input magnetograms, which serve as the bottom boundary condition of the extrapolation. We corrected the projection effects and removed the Lorentz force and torque using the method of \citet{Wiegelmann2006} so as to conform to the force-free assumption. 
This approach entails an optimization method to minimize the function composed of four terms \citep[Equation (6) in][] {Wiegelmann2006}, including the deviation from the observational data, smoothness of the magnetic fields, and the Lorentz force and torque. By taking this approach, the Lorentz force and torque can be  reduced to a small value, resulting in a smooth vector magnetic field within the measurement error. We performed 5000 iterations for each magnetogram, after which the Lorentz force and torque decreased to one-thousandth of the original magnitudes \citep[quantified by $\epsilon$$_{force}$ and $\epsilon$$_{torque}$ in][] {Wiegelmann2006}. We can state that the forces are nearly null and the magnetic configuration is almost stable.
Subsequently, we reconstructed the 3D coronal magnetic fields. As shown in the observations, the eruptive filament is located at the edge of the active region, which can be categorized as intermediate type. In this case, it was difficult to construct the sheared and twisted magnetic fields of the filament for two reasons. On the one hand, the vector magnetic fields in decayed regions are highly noisy in general. On the other hand, the intermediate filament is generally high-lying, meaning that the information is harder to transform precisely from the input photosphere magnetograms to the positions of the filament due to the errors during the numerical computation. To this end, following the method of our previous works \citep{Guo2019,Guo2021}, we embedded a twisted flux rope with the regularized Biot-Savart Laws (RBSL) method into the potential field, in which the axis of the inserted flux rope is parellel to the path of the filament in observations. Next, we relaxed the aforementioned magnetic fields to a force-free state with the magneto-frictional model. After 60000 iterations, the force-free and divergence metrics are 0.35 and $3.55 \times 10^{-5}$, respectively. Compared to our previous studies, the values of these two metrics are acceptable.

Figure~\ref{NLFFF} shows the results of the NLFFF extrapolation, which are overlaid on the viewing angle of AIA observations. The green, pink, and dark cyan tubes represent the magnetic structures of the eruptive filament, the fan-spine structure, and the peripheral active regions, respectively. Based on the alignment between the typical field lines and flare ribbons, we conclude that two parallel and circular flare ribbons are formed due to magnetic reconnection in overlying arcades and the fan-spine reconnection. Additionally, we also find that the dimming is surrounded by circular ribbons (Figure~\ref{eruption}), which may suggest an interchange magnetic reconnection is taking place, which could be causing the jump and expansion of the flux rope footpoint. In conclusion, the NLFFF extrapolation at the scale of the active region explains many physical processes seen in observations, and in particular the circular ribbons and dimmings.   The inserted flux rope visualized in Figure \ref{NLFFF}  represents the filament before its eruption and subsequent change of direction.  The inserted RBSL flux rope represents  an extended  filament adopting the global magnetohydrodynamic (MHD) modeling.
The global view of the ribbons in Figure \ref{eruption} corresponds to this configuration.   It is clear that the remote circular  brightening and dimming, labeled  S2 in Fig. 2, corresponds to this complex nest of active regions in a multipolar configuration.  The remote circular brightening and dimming could be related to footpoints of arcades corresponding to the fan-spine.
A data-driven analysis should be performed to  clearly  understand the process of rotation and deflection of the filament before the  ejection of the CME.

The Type \uppercase\expandafter{\romannumeral3} radio burst (Figure~\ref{radio}) indicates an interchange reconnection with the open field lines. To verify the existence of open field lines, we performed a global MHD modeling including the solar wind using the model COCONUT (\citetalias{PerriLeitner2022}~\citeyear{PerriLeitner2022}). Based on this global MHD modeling, we were able to investigate the magnetic structures near this active region. This modeling includes the following steps: First, the initial magnetic fields are provided by the potential-field source-surface (PFSS) model computed with a fast finite-volume solver in COCONUT (\citetalias{PerriLeitner2022}~\citeyear{PerriLeitner2022}). The bottom input magenetogram is preprocessed with the spherical harmonics projection with a maximum frequency of $l_{\rm max}=30$. Hereafter, we relax the above state with the polytropic MHD model. After 35700 relaxation steps, a convergence level of $10^{-5}$ is reached, meaning that the quasi-static solar-wind solution has been attained (see \citet{GuoJH2024}). 

Figure~\ref{global_corona} shows the typical field lines in the global coronal modeling. One can see fan-spine structure (green tubes), overlying closed arcades (purple tubes), and the open field lines (yellow tubes) 
toward the south, corresponding to the coronal hole (panel a). We posit that the magnetic reconnection between the eruptive flux rope and the open fields can inject energetic particles into the interplanetary space, driving a type III radio burst.

\section{Discussion and conclusions}
In this paper, we present a detailed analysis of a filament eruption observed by SDO, STEREO, SUTRI, and GONG simultaneously.  There are two stages to the filament eruption: in the first, the filament rises up toward the northeast, and in the second, the direction of the eruption is deflected toward the southwest, at which point the eruption  can be considered nonradial. 

Based on the combination of simultaneous observations from multiple instruments, NLFFF extrapolation, and a polytropic global MHD modeling, we propose the following scenario: In the beginning, the filament is initially restrained below a dome of a fan-spine configuration (shown by the pink lines in Fig. \ref{NLFFF}). Due to the continuous magnetic activities at the footpoints, the filament  gradually begins to rise toward the northeast. According to \citet{Aulanier2019}, it is possible to define the geometry of three types of 3D magnetic reconnections, namely ``aa-rf'' reconnection, ``ar-rf'' reconnection, and ``rr-rf'' reconnection. In our case, it is possible we are seeing an aa-rf magnetic reconnection above the filament, resulting in two bright flare ribbons and post-flare loops. Such reconnection can disassemble the restraint above the filament and drive the filament to keep rising, which sets the stage for the magnetic reconnection later on. One of the flare ribbons expands along a circular trajectory. When the filament reaches a certain height, it reconnects with the open field lines besides it (yellow lines in Figure \ref{global_corona} ).  Plasma is accelerated at the reconnection point and propagates outward along the open magnetic lines, and so the plasma density decreases and the region darkens. After the reconnection, some new loops form. It is worth noting that the direction of the filament later in eruption is guided by the open magnetic field lines to the east. The direction of its deflection is the same as the direction in which the open magnetic field lines bend \citep{Yang2015,Yang2018}. This corresponds to subsequent observations of the CME.

This example provides evidence for the multiple magnetic reconnections experienced by large -scale filaments during eruption. Both the background magnetic field and the filament fibrils and strands  contribute during magnetic reconnection. The findings presented here providing a new understanding of the diverse magnetic fields and energy exchanges of solar eruptive activity.  We look forward to developing more high-resolution instruments in the future and using high-resolution data to study the filament-eruption process in as much detail as possible.
\begin{acknowledgements}
We thank the referee for careful reading and many constructive comments. This work is supported by the NSFC grant 12173022, 11790303 and 11973031. The authors are indebted to the SDO, STEREO, SUTRI, GONG, SOHO and CALLISTO teams for providing the data.  We are very grateful to Zhang Liang for his help in data processing. SUTRI is a collaborative project conducted by the National Astronomical Observatories of CAS, Peking University, Tongji University, Xi'an Institute of Optics and Precision Mechanics of CAS and the Innovation Academy for Microsatellites of CAS. Data were acquired by GONG instruments operated by NISP/NSO/AURA/NSF with contribution from NOAA. SOHO is a project of international cooperation between ESA and NASA.
\end{acknowledgements}

\bibliography{20221004}
\bibliographystyle{aa}

\begin{figure*}[!htbp]
\begin{minipage}{\textwidth}
\centering
\includegraphics[width=160mm,angle=0,clip]{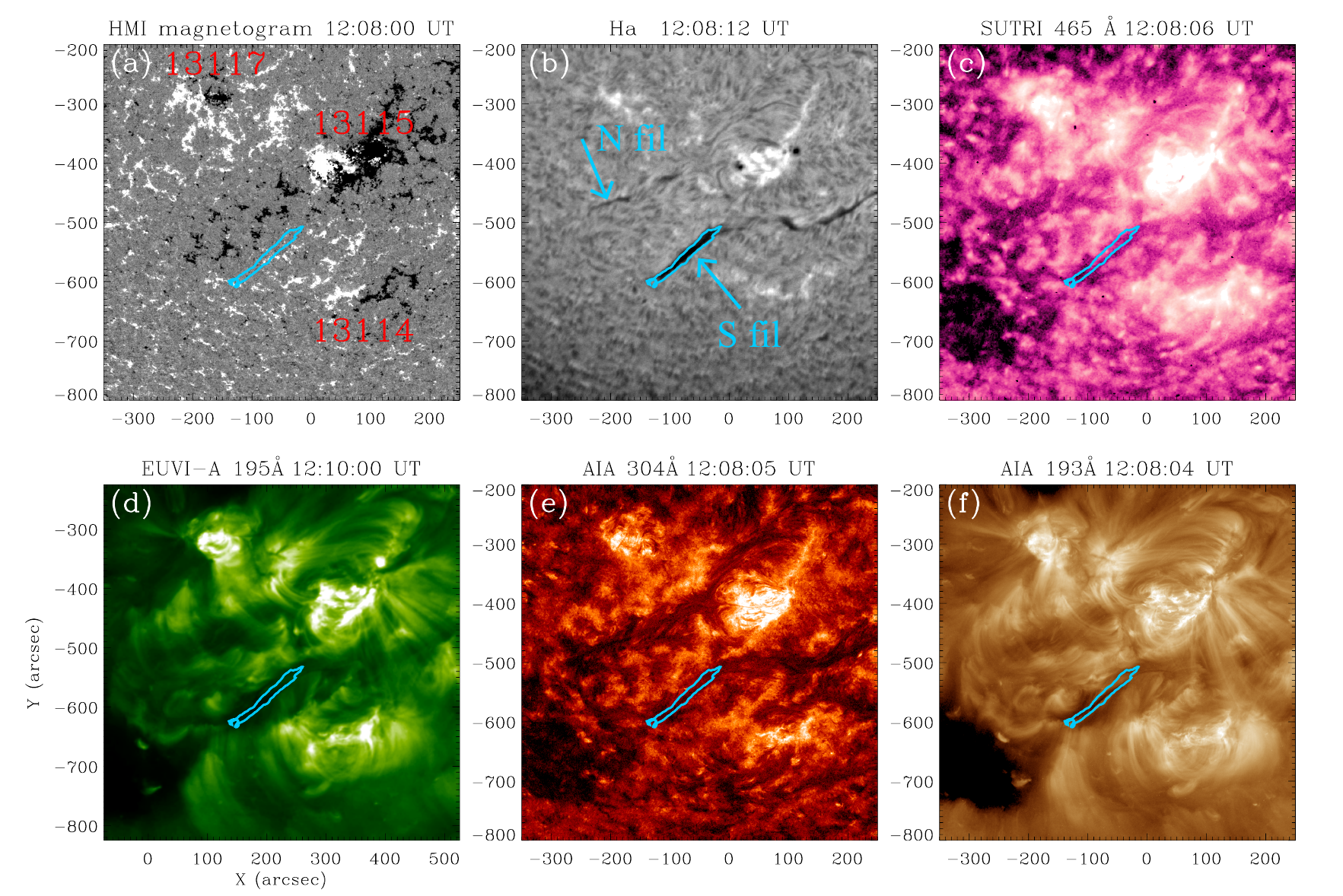}
\end{minipage}
\caption{ Overview of  the filament  in a nest of three active regions. The northern  filament (Nfil) does not evolve while the southern filament (Sfil) erupts, i.e., mainly its eastern part, which is marked with cyan contours in all the panels. 
The images were  obtained by  SDO/HMI, SDO/AIA, GONG, SUTRI, and STEREO/EUVI-A on 2022 October 4 before eruption. The filament is visible from the chromosphere to the transition region and the corona as a dark channel. The magnetogram shows that the filament lies  in a mixed-polarity area between AR 13115 and AR 13114  (panel a)  and extends along  350 arcsec in longitude (panel b). 
}
\label{position}
\end{figure*}
\begin{figure*}[!htbp]
\begin{minipage}{\textwidth}
\centering
\includegraphics[width=160mm,angle=0,clip]{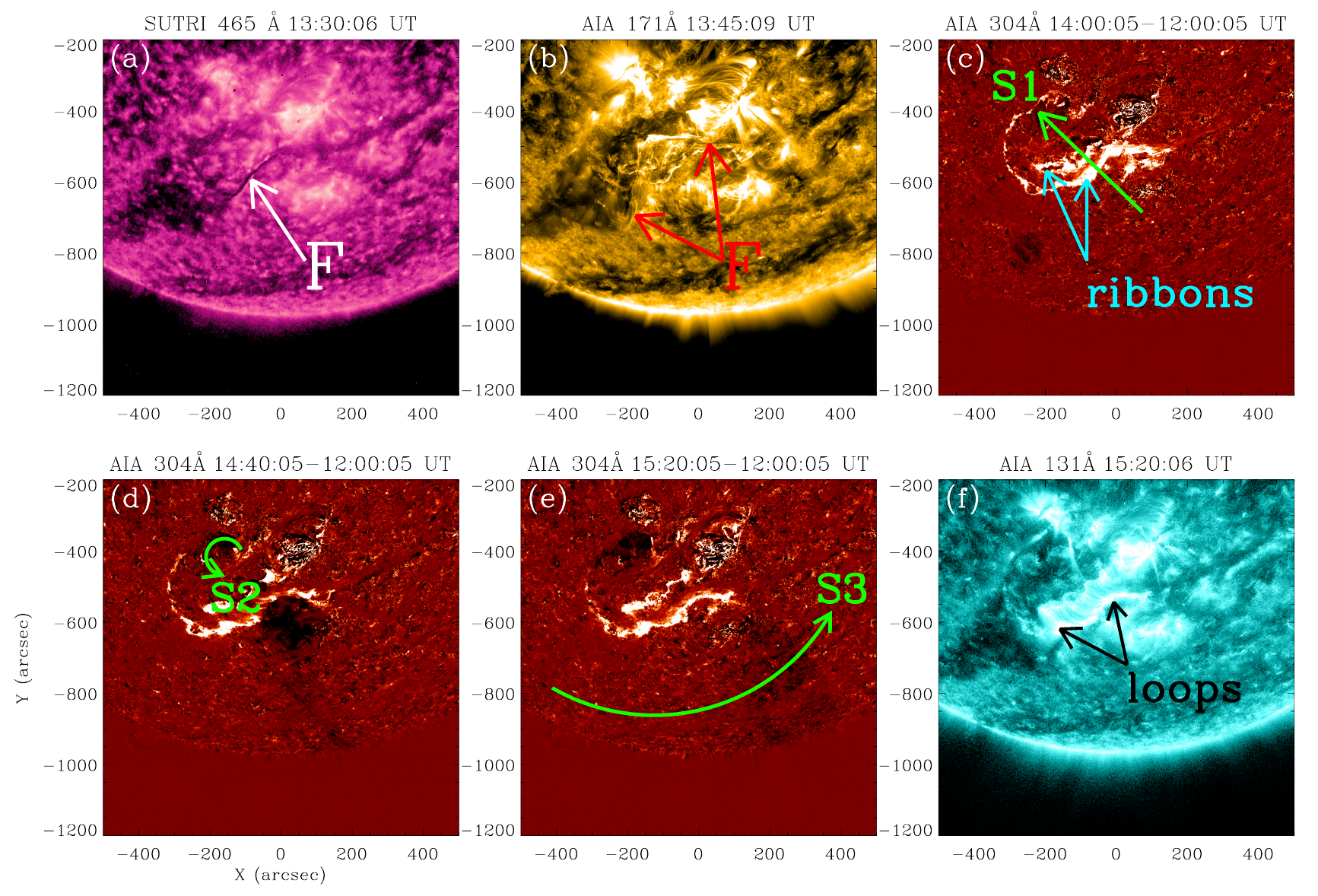}
\end{minipage}
\caption{Evolution of the filament eruption in multiple wavelengths   between 13:30 UT and  15:20 UT. The filament rises up  (white and red arrows  in panels a and b).  Two flare ribbons  are highly visible in difference images in AIA 304 \AA\ (14:00-14:40-15:20  minus 12:00 UT) (panels c, d, and e).  Cyan and black arrows stand for flare ribbons and post-flare loops in panels c and f,  respectively. At 14:00 UT, the south ribbon is extended and turns to  the north, creating     a remote brightening. At its end, a remote dimming  appears at 14:00 UT in S2   during the eruption. Three slices, labeled S1, S2, and S3, along the direction of filament eruption (initial stage), the dimming region propagation, and filament deflection are indicated by green arrows (panels c, d,  and e). S1 is along the straight arrow, and S2 and S3 are the arc slices. These slices are used  for the time--distance analysis in Figures \ref{304slice} and \ref{dimming}. }
\label{eruption}
\end{figure*}
\begin{figure}[!htbp]
\resizebox{\hsize}{!}{\includegraphics[width=120mm,angle=0,clip]{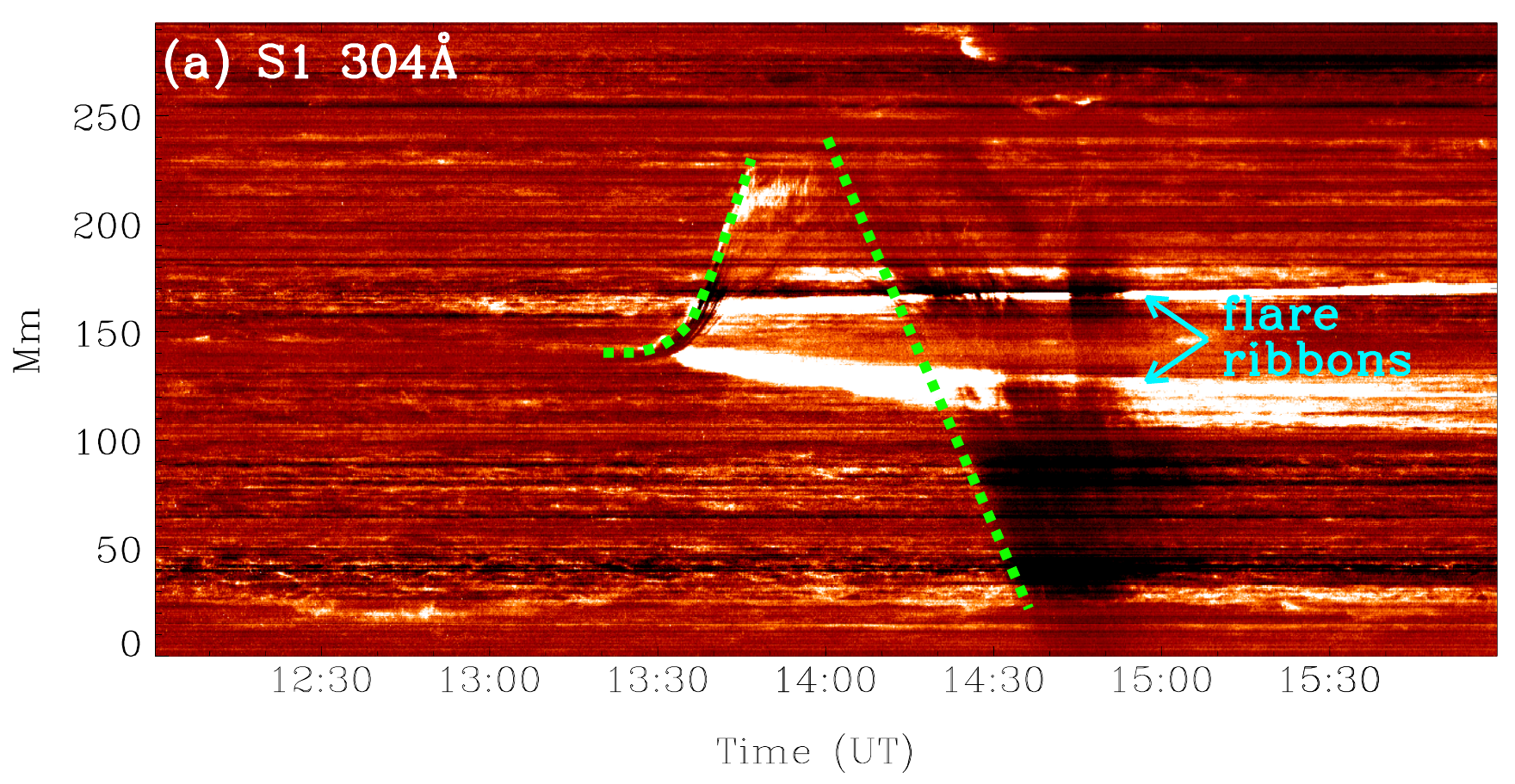}}
\resizebox{\hsize}{!}{\includegraphics[width=120mm,angle=0,clip]{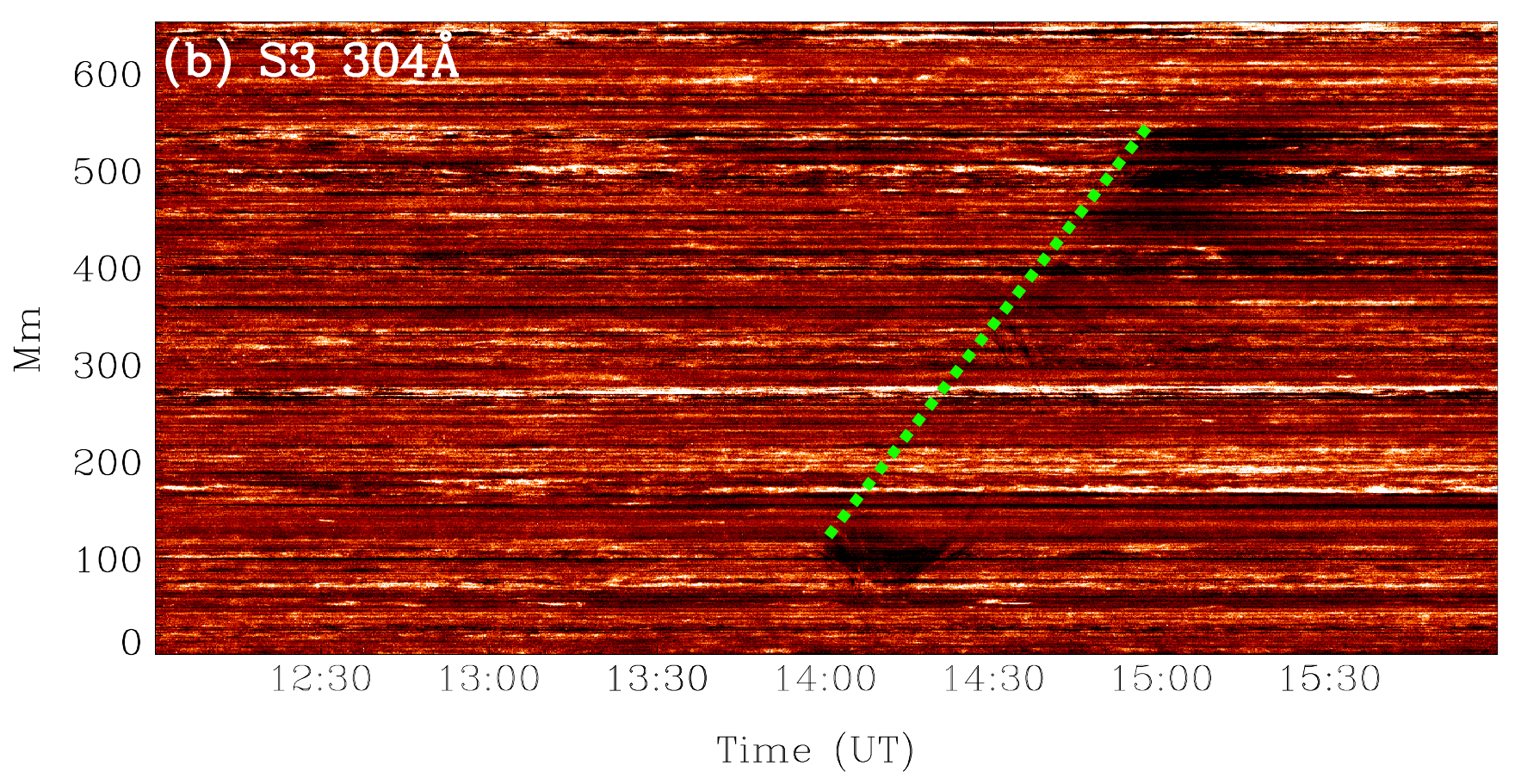}}
\caption{Time--distance diagrams of  the filament in AIA   304\AA{} for two stages. The diagrams show the rising and initial eruption of the filament  along S1 (top panel) and subsequent deflection (bottom panel) along S3. The positions of flare ribbons are marked by cyan arrows.}
\label{304slice}
\end{figure}
\begin{figure*}[!htbp]
\begin{minipage}{\textwidth}
\centering
\includegraphics[width=90mm,angle=0,clip]{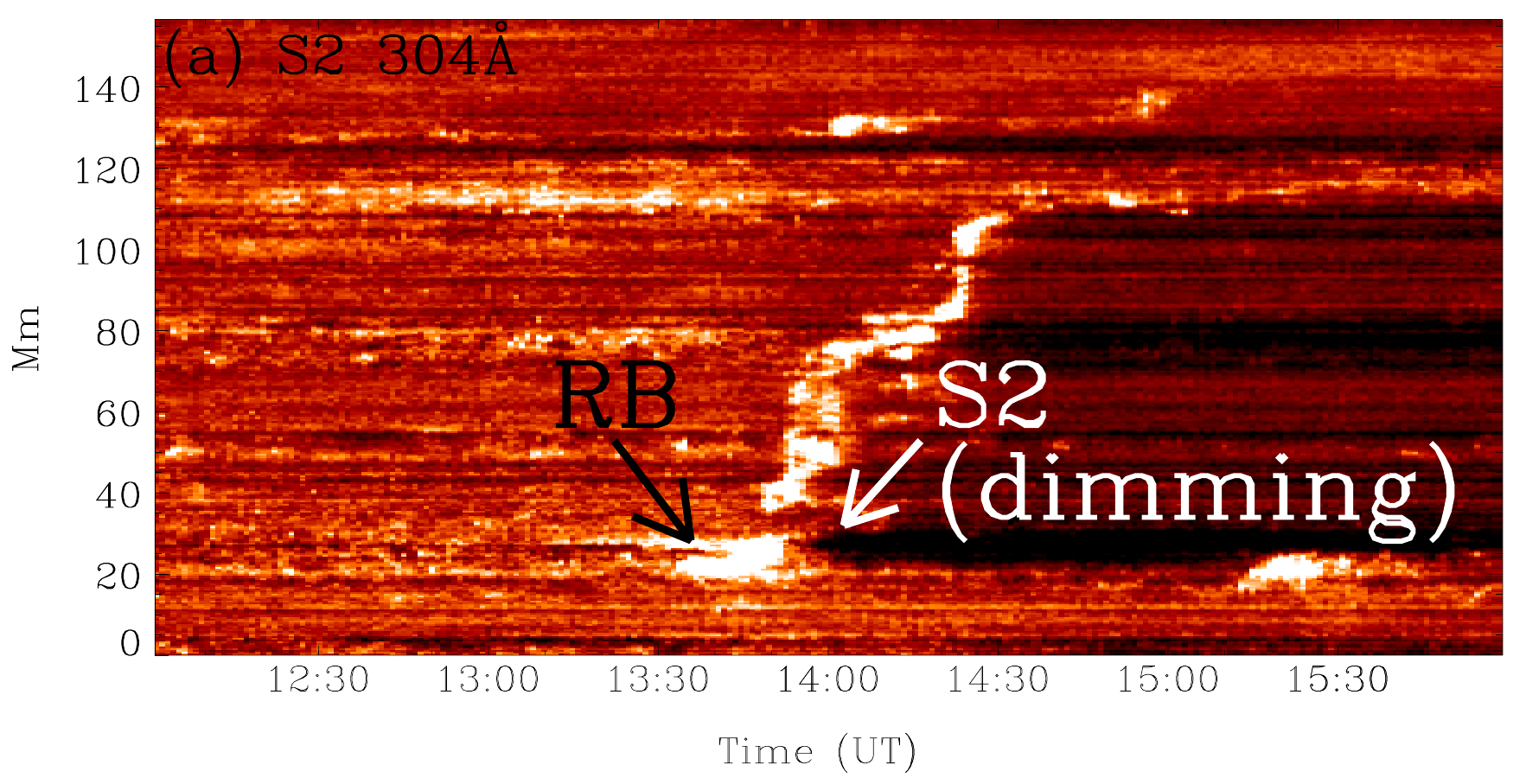}
\includegraphics[width=90mm,angle=0,clip]{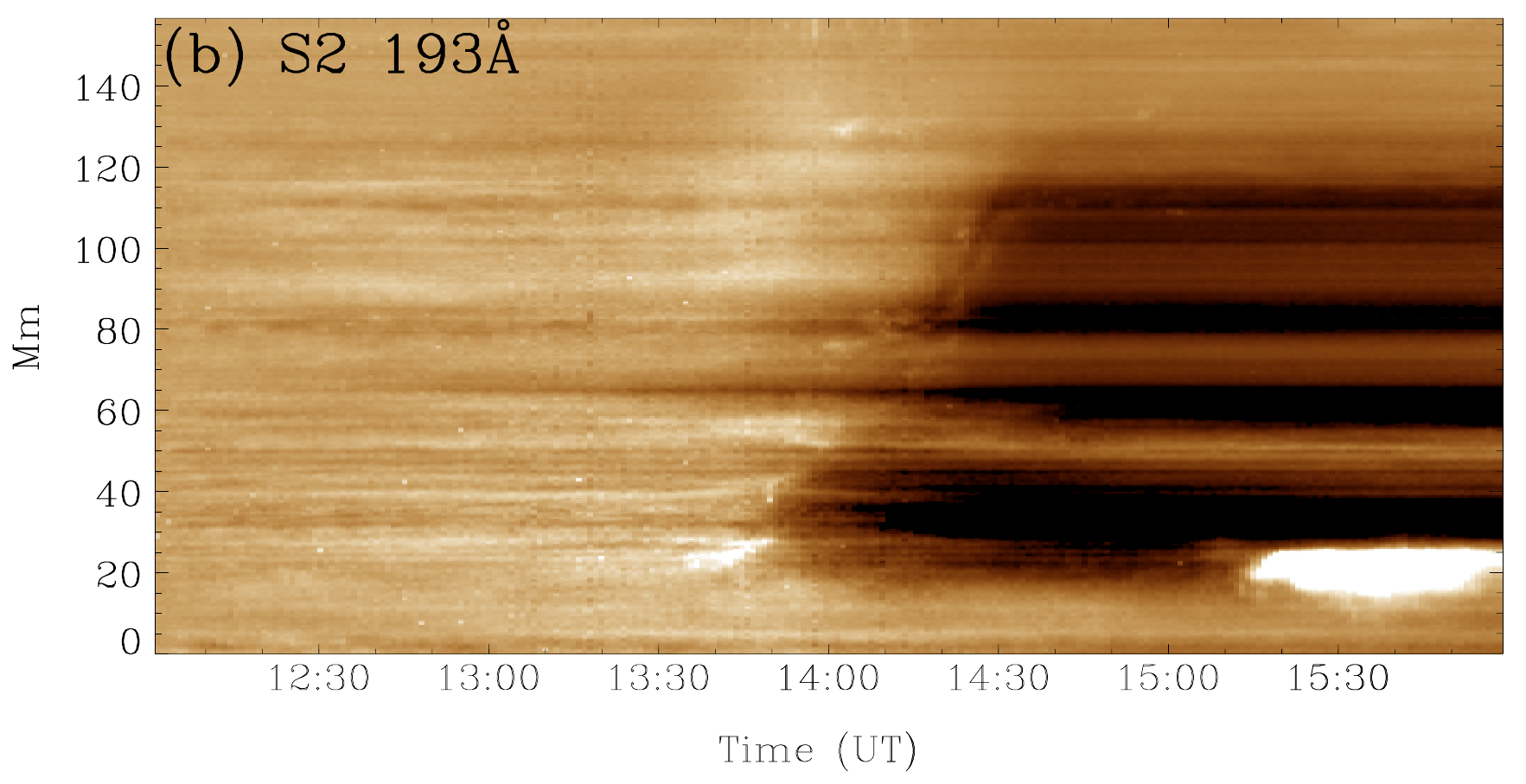}
\includegraphics[width=90mm,angle=0,clip]{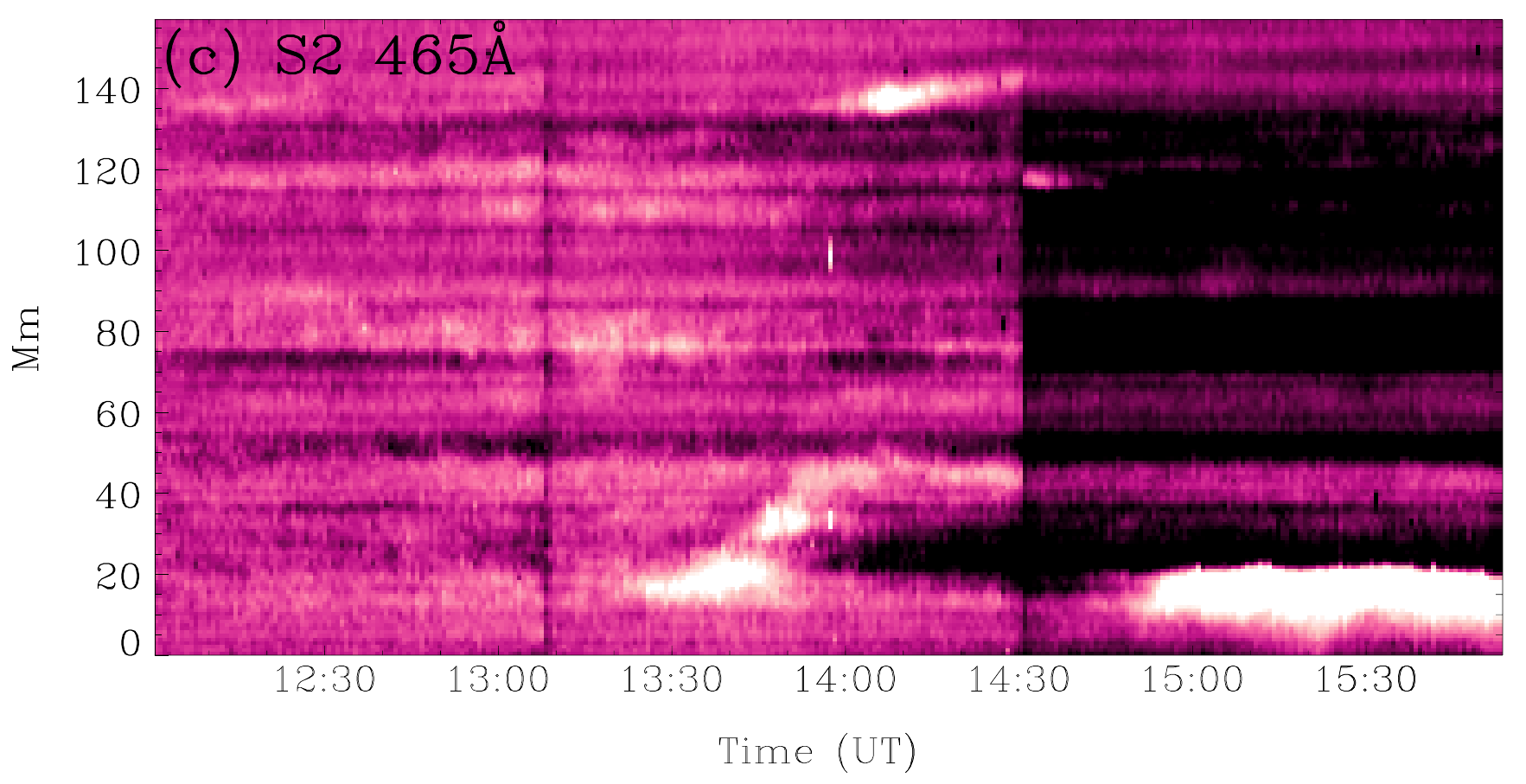}
\includegraphics[width=90mm,angle=0,clip]{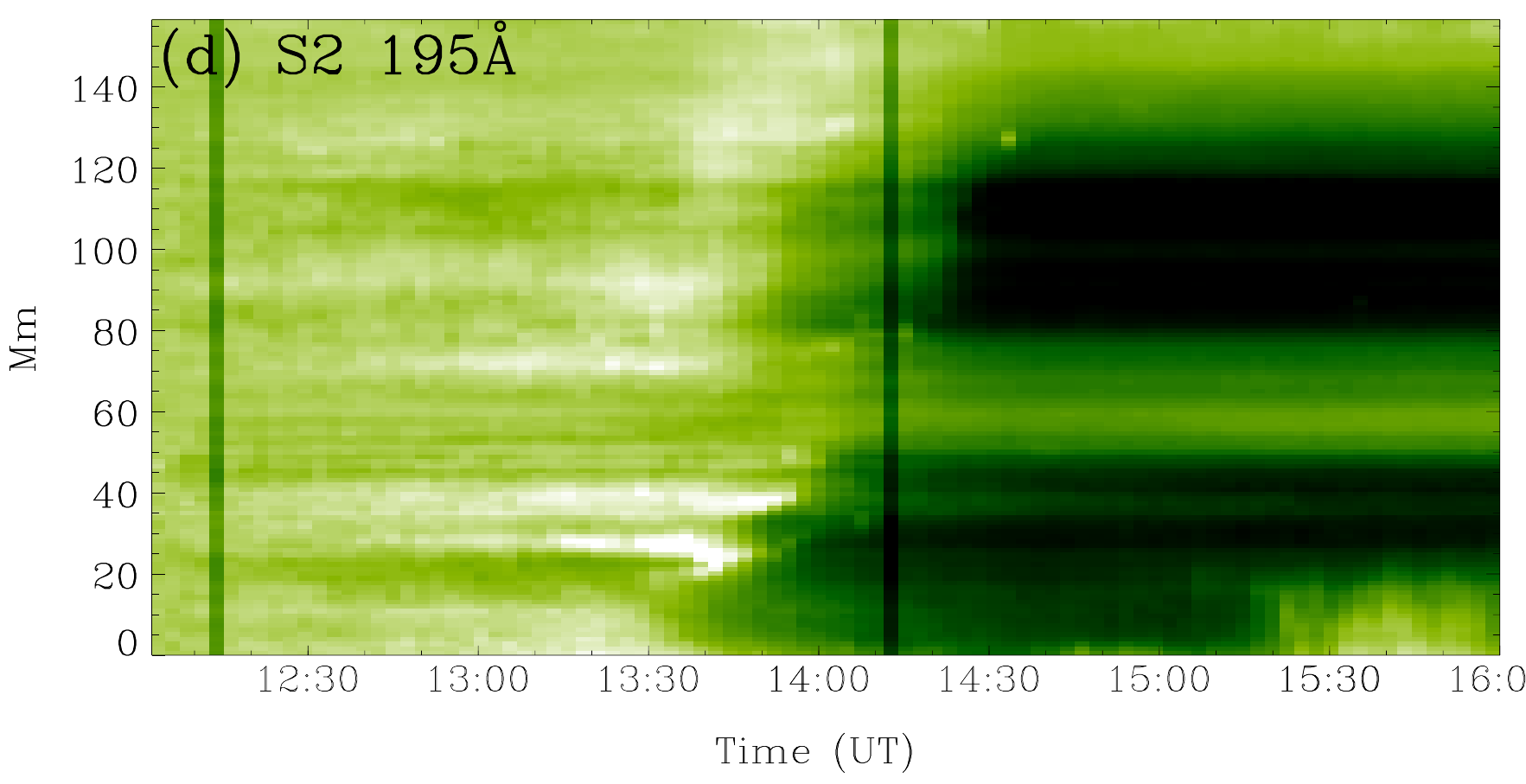}
\end{minipage}
\caption{Time--distance diagrams of the dimming region in EUV bands of S2. Black and white arrows represent the remote brightening (labeled `RB') and remote dimming  region (S2), respectively. }
\label{dimming}
\end{figure*}
\begin{figure*}[!htbp]
\begin{minipage}{0.45\textwidth}
\centering
\includegraphics[width=80mm,angle=0,clip]{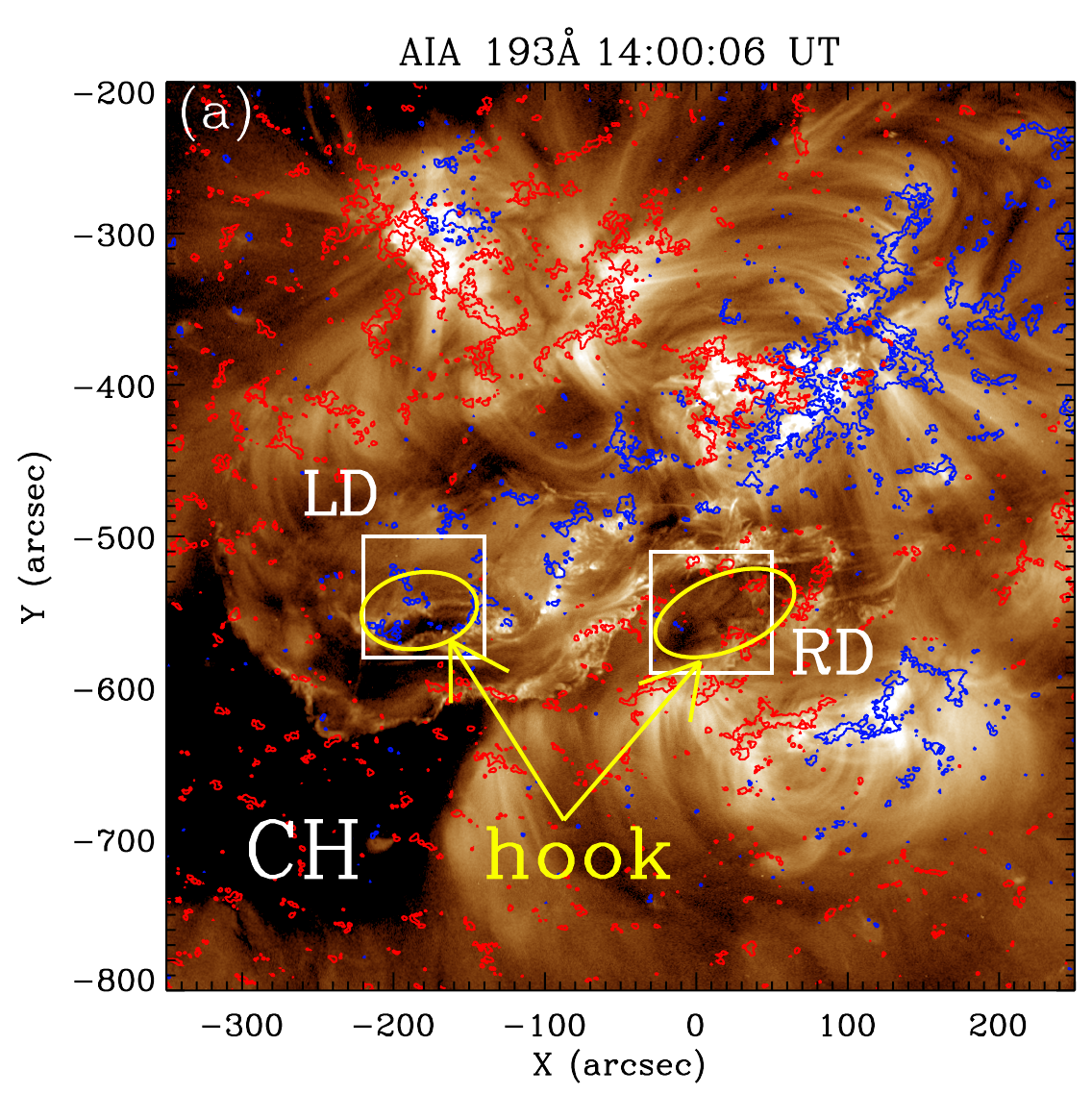}
\end{minipage}
\begin{minipage}{0.5\textwidth}
\centering
\includegraphics[width=80mm,angle=0,clip]{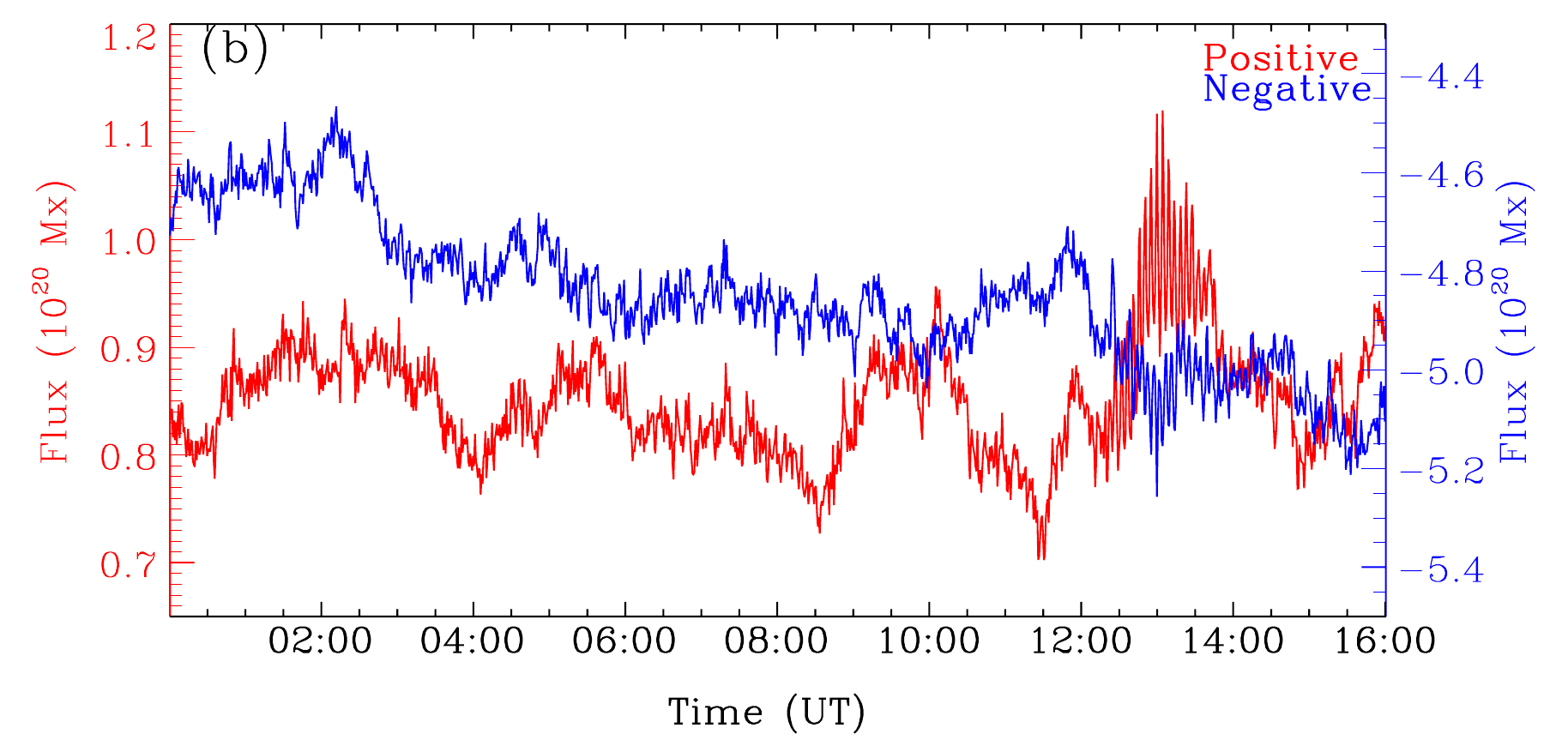}
\includegraphics[width=80mm,angle=0,clip]{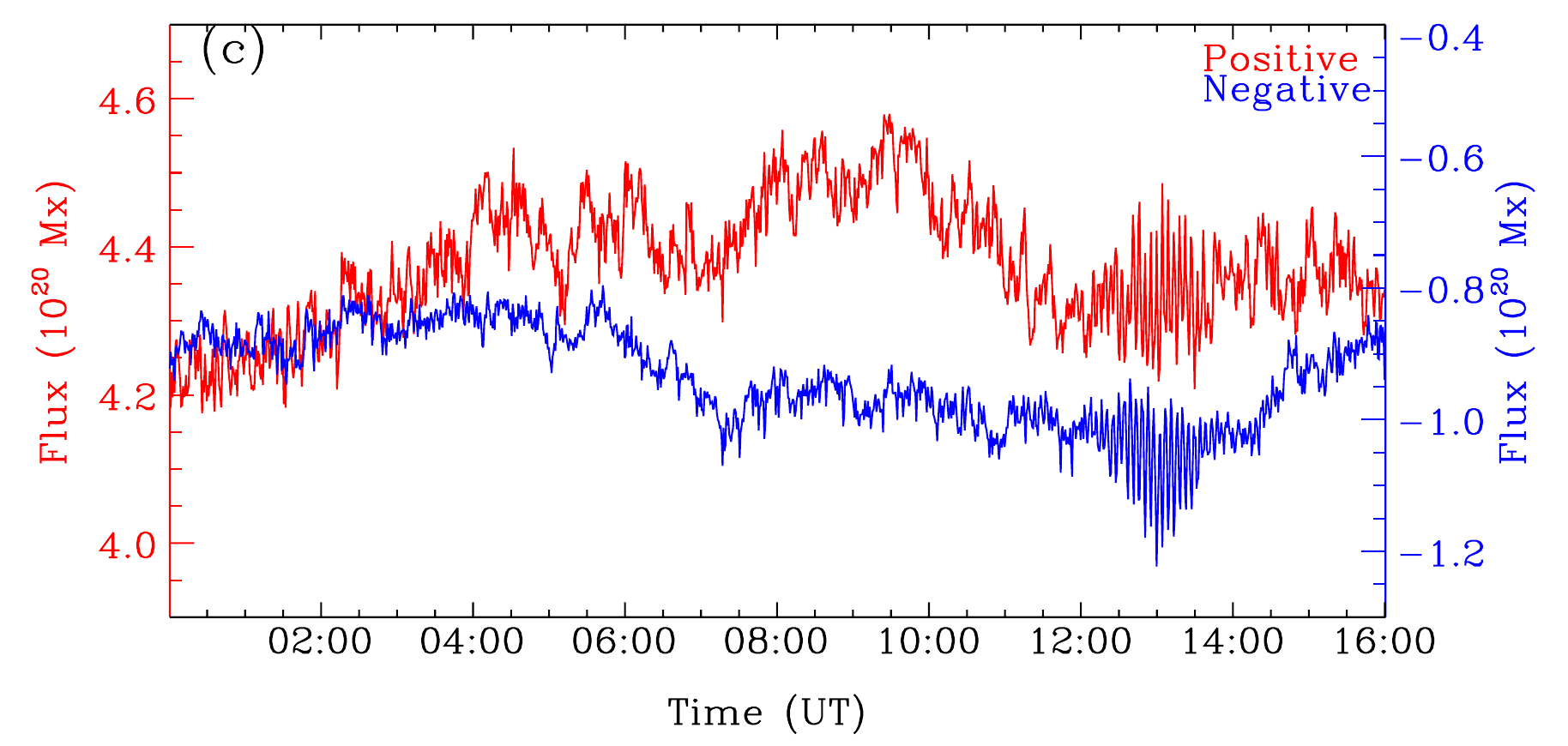}
\end{minipage}
\caption{Evolution of magnetic flux at  two filament footpoints. The 193 \AA{} image is overlaid with the longitudinal magnetic field. Red and blue contours indicate the magnetic field Bz at -50G and 50G, respectively. The white boxes represent the areas where the flux is calculated.  The evolution of the left dimming (LD) flux is shown in panel (b) while the right dimming (RD) is shown in panel (c). CH   (written in  white)  represents  the  coronal hole. Yellow ellipses and arrows indicate the hooks at the ends of the flare ribbons. }  
\label{flux}
\end{figure*}
\begin{figure*}[!htbp]
\begin{minipage}{\textwidth}
\centering
\includegraphics[width=160mm,angle=0,clip]{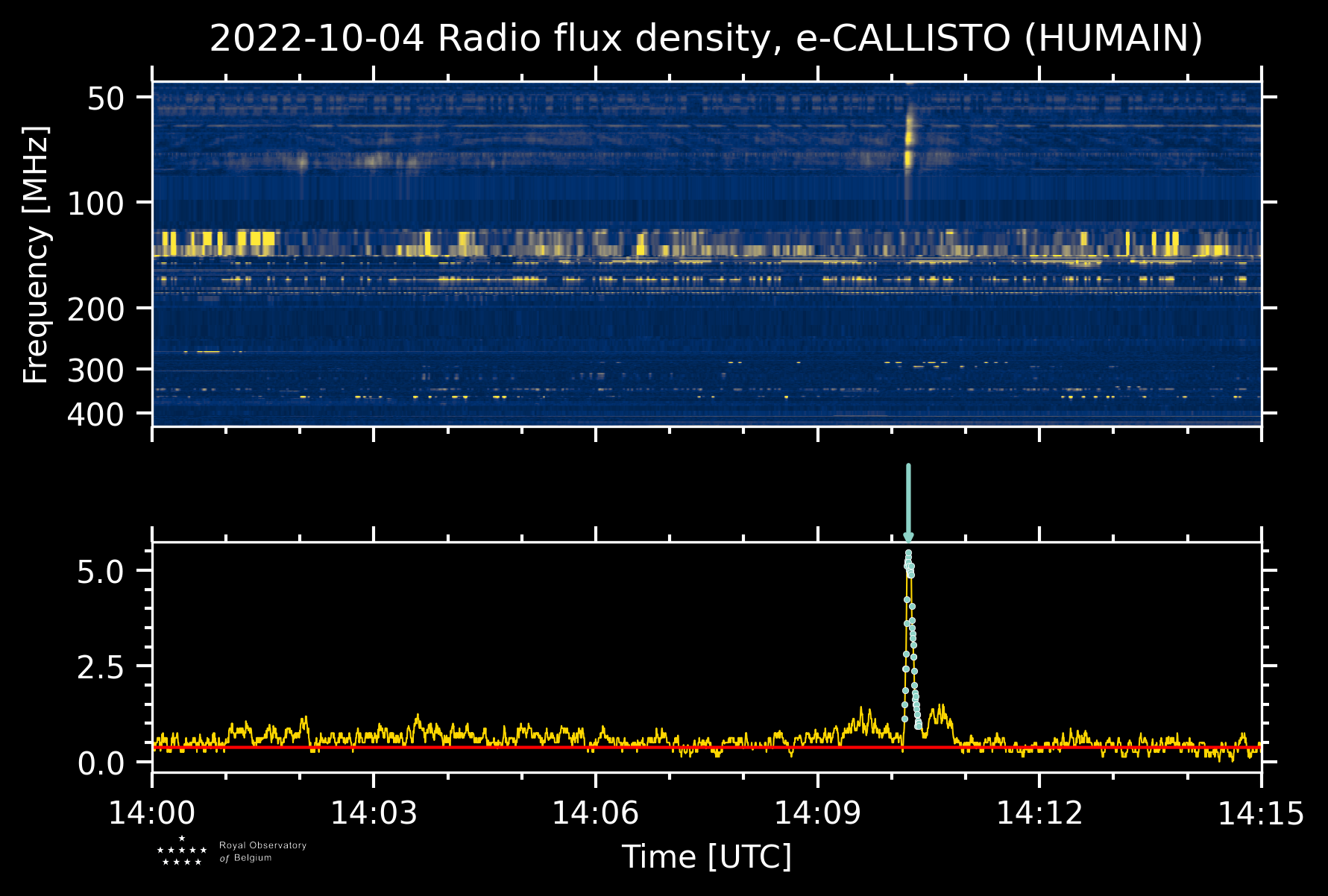}
\end{minipage}
\caption{Radio dynamic spectra from
CALLISTO, exhibiting  a Type \uppercase\expandafter{\romannumeral3} radio burst (cyan peak). }
\label{radio}
\end{figure*}
\begin{figure*}[!htbp]
\begin{minipage}{\textwidth}
\centering
\includegraphics[width=160mm,angle=0,clip]{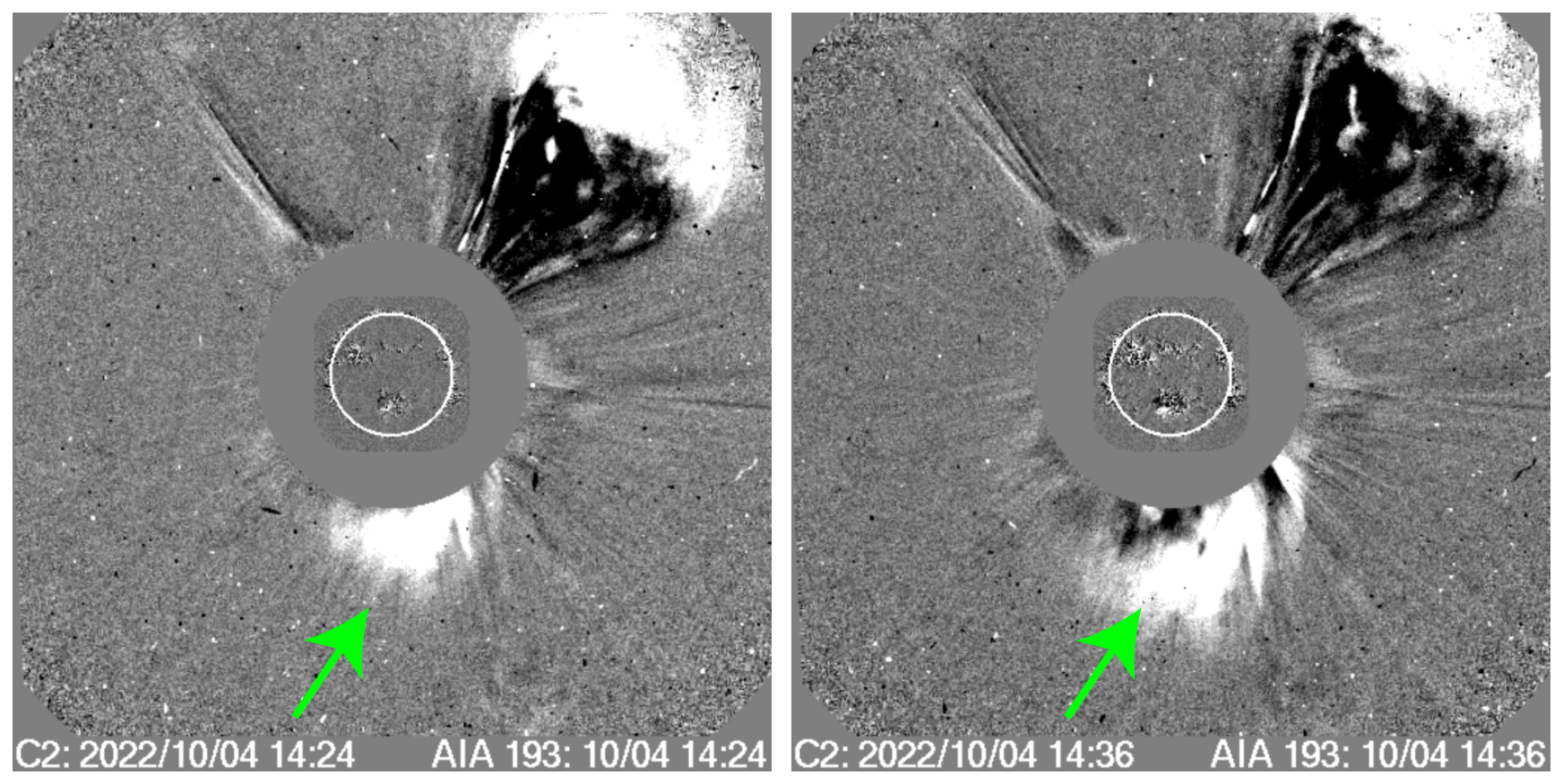}
\end{minipage}
\caption{Coronal mass ejections   observed by LASCO. The CME to the south is  correlated with the  filament eruption 
marked by  a green arrow.  The north--west CME  is ejected  at 13:25  UT from another AR located close to  the western limb  (AR 13113). } 
\label{cme}
\end{figure*}
\begin{figure}[!htbp]
\resizebox{\hsize}{!}{\includegraphics[width=120mm,angle=0,clip]{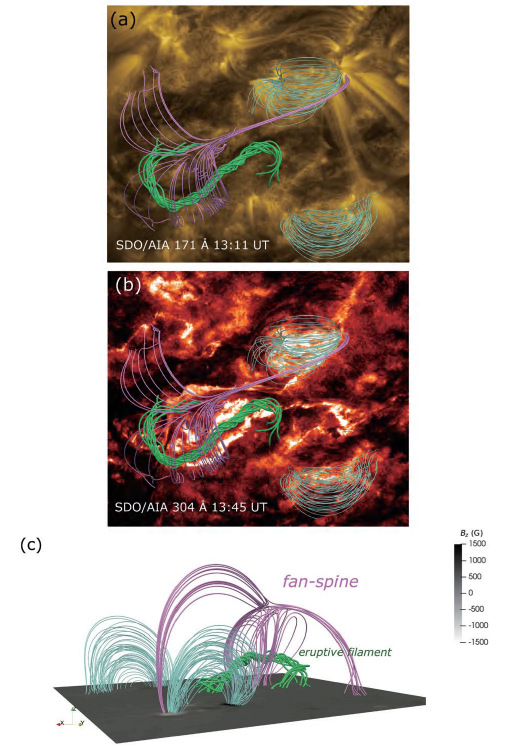}}
\caption{Magnetic configuration of the three  ARs (13114, 13115, 13117) and the intermediate filament  before  activity provided by a NLFFF magnetic field extrapolation using a  magneto-frictional model. Panels (a) and (b)  show magnetic field lines presented from the perspective of AIA observations, and panel (c) shows these in a 3D view. The green twisted field lines represent the regularized Biot-Savart flux rope (eruptive filament) inserted in the magnetic extrapolated configuration, the dark cyan  lines represent  the overlying arcades in the AR and its surroundings, and the pink lines represent  the fan and the spine over the filament.  }
\label{NLFFF}
\end{figure}
\begin{figure*}[!htbp]
\begin{minipage}{\textwidth}
\centering
\includegraphics[width=160mm,angle=0,clip]{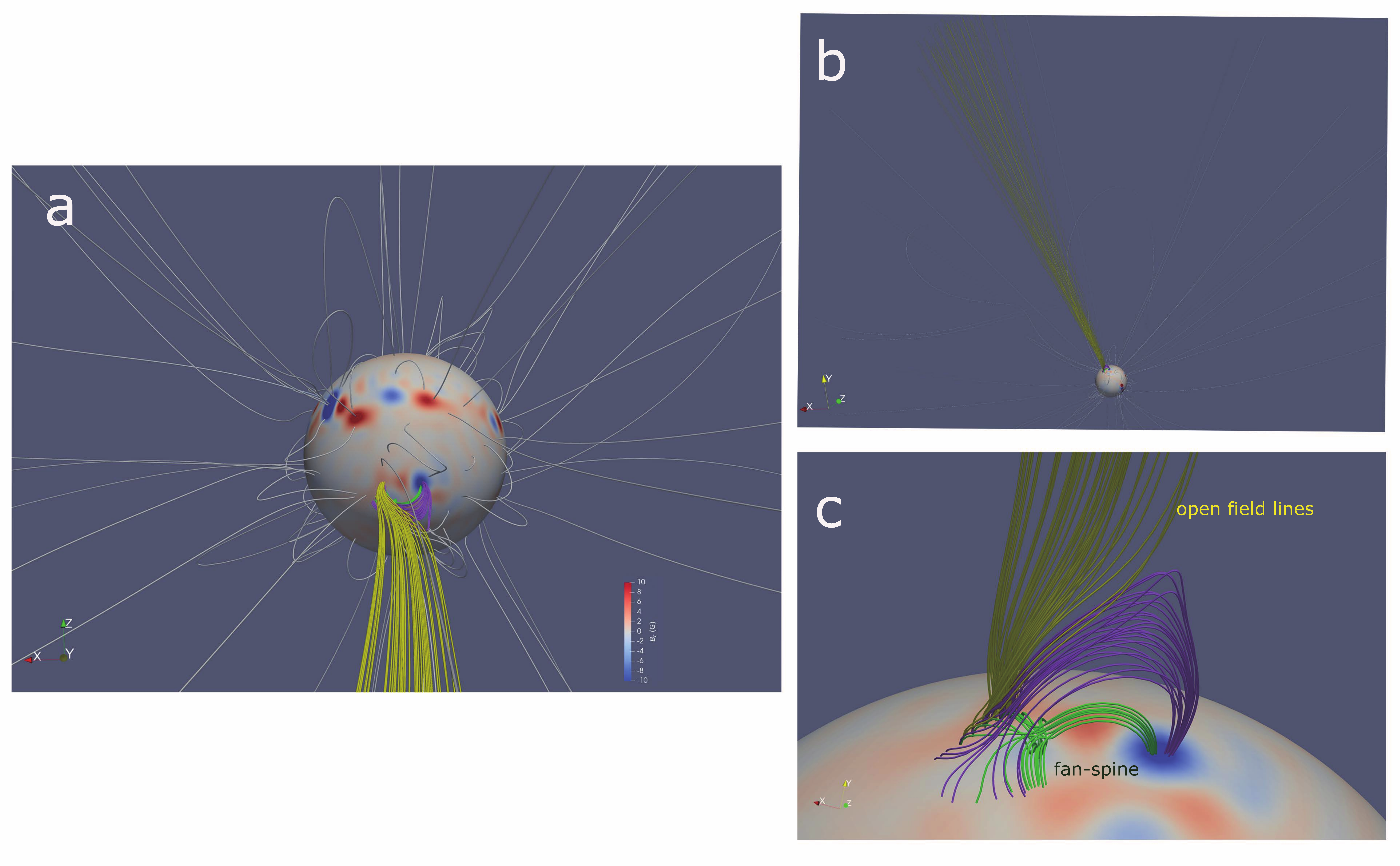}
\end{minipage}
\caption{Global coronal
modeling with COCONUT of the AR environment,   showing the reconnection with open field lines in the region of the dimming. The fan-spine structure (green tubes), overlying closed arcades (purple tubes), and the open field lines (yellow tubes) are  presented.  Panels (b) and (c) have been rotated by 180 degrees to show the open field lines. }
\label{global_corona}
\end{figure*}


\end{document}